\documentclass[a4paper]{jpconf}
\usepackage{graphicx}
\usepackage{amsmath}
\newcommand{\hc}{\hbox {h.c.}}

\begin{document}
\begin{flushright}
CERN-TH-2017-070
\end{flushright}
\title{Spontaneous symmetry breaking in three-Higgs-doublet $S_3$-symmetric models}

\author{D Emmanuel-Costa$^{1,}$\footnote[5]{Affiliation valid till the end of 2016}   
O M Ogreid$^2$, P Osland$^3$ and 
\underline{M N Rebelo}$^{1,}$$^4$}

\address{$^1$ Centro de F\'isica Te\'orica de Part\'iculas -- CFTP and Dept de F\' \i sica
Instituto Superior T\'ecnico -- IST, Universidade de Lisboa (UL), Av. Rovisco Pais,
P-1049-001 Lisboa, Portugal}
\address{$^2$ Western Norway University of Applied Sciences,
Postboks 7030, N-5020 Bergen, Norway } 
\address{$^3$ Department of Physics and Technology, University of Bergen, 
Postboks 7803, N-5020  Bergen, Norway} 
\address{$^4$  Theoretical Physics Department, CERN, CH-1211, Geneva 23, Switzerland}

\ead{omo@hvl.no, Per.Osland@uib.no and rebelo@tecnico.ulisboa.pt}

\begin{abstract}
The talk summarises work done by the authors consisting of a detailed study of
the possible vacua in models with three Higgs doublets 
with $S_3$ symmetry and without explicit CP violation. Different vacua require
special regions of the parameter space which were analysed in our work.
We establish the possibility of spontaneous CP violation in
this framework and we also show which complex vacua conserve CP.
In our work we discussed constraints from vacuum stability. 
The results presented here are relevant for model building.
\end{abstract}

\section{Introduction}
In the Standard Model of Particle Physics (SM) there is one Higgs doublet 
responsible for spontaneous electroweak symmetry breaking and for the 
mechanism that gives mass to fermions and to electroweak gauge bosons.
The model predicts the existence of one Higgs boson. In 2012 a scalar boson 
was discovered at the LHC \cite{Aad:2012tfa,Chatrchyan:2012xdj}
with properties consistent with those predicted
by the SM. However there are good motivations to consider models with more
than one Higgs doublet such as the possibility of having CP symmetry broken
spontaneously \cite{Lee:1973iz} or new sources of CP violation. Supersymmetric models
require two Higgs doublets. Furthermore, models with two Higgs 
doublets have a rich phenomenology with many interesting possible 
manifestations of physics beyond the SM \cite{Gunion:1989we,Branco:2011iw}. 
Extensions of the SM with more than one  Higgs doublet are good
candidates to explain some of the present flavour anomalies and to 
solve some of the puzzles left unanswered by the SM. 

Models with more than one Higgs doublet can give substantial contributions
to flavour changing neutral currents (FCNC). Current experimental bounds 
require these to be strongly suppressed. One possibility is to completely 
forbid Higgs mediated FCNC at tree level via a symmetry, as is the case
in models with natural flavour conservation (NFC) \cite{Glashow:1976nt, 
Paschos:1976ay} where only one Higgs 
doublet is allowed to couple to each charge quark sector. In the 
case of two Higgs doublets this is achieved by means of a $Z_2$
symmetry and as a result neither spontaneous nor hard CP violation can 
occur in the Higgs sector. It is possible to have CP
violation in the scalar sector, with two Higgs doublets and NFC, with the $Z_2$
symmetry softly broken in the Higgs potential \cite{Branco:1985aq}.
Three Higgs doublets and NFC with exact $Z_2$ symmetries allow for CP to be 
violated either explicitly \cite{Weinberg:1976hu}
or spontaneously \cite{Branco:1980sz}.  FCNC can also be eliminated at
tree level assuming alignment of Yukawa couplings in flavour space \cite{Pich:2009sp},
in this case complex Yukawa couplings give rise to new sources of CP violation. 
There have been several attempts at obtaining Yukawa alignment in several
extensions of the SM  \cite{Serodio:2011hg,Varzielas:2011jr, 
Celis:2014zaa, Botella:2015yfa}.
Another very interesting possibility 
to avoid the problem of having too large FCNC with two Higgs doublets, 
is to build models allowing for Higgs mediated FCNC which are under control 
by means of a symmetry that leads to couplings suppressed by small 
off-diagonal elements of the quark mixing matrix $V_\text{CKM}$. The first 
models of this type, based on a symmetry, were proposed by Branco, Grimus and 
Lavoura (BGL) \cite{Branco:1996bq}. 
Implications of BGL-type models have been extensively analysed 
recently, in the light of the LHC results, for several different 
implementations \cite{Botella:2009pq,Botella:2011ne, 
Bhattacharyya:2013rya, Botella:2014ska, Bhattacharyya:2014nja,
Botella:2015hoa}.
BGL-type models can have FCNC either in the up sector or in the down 
sector but not in both sectors at the same time. Recently, a generalisation
of BGL-type models allowing for FCNC in both sectors was 
built \cite{Alves:2017xmk}.
BGL models models can also be extended to the case of three 
doublets \cite{Botella:2009pq}.

Three-Higgs-doublet models may provide good dark matter 
candidates \cite{Grzadkowski:2009bt,Grzadkowski:2010au,Keus:2014jha,Machado:2012ed,
Fortes:2014dca,Fortes:2014dia}. 
One may also speculate that nature is such that three generations of fermions 
come with three Higgs doublets, which is, of course, an issue to be settled
by experiment. As the number of doublets increases so does 
the complexity of the scalar potential and the  number
of free parameters in the theory \cite{Olaussen:2010aq}. 
Discrete symmetries play an important role in reducing
this number and lead at the same time to testable predictions. Symmetries
also play an important role in stabilising dark matter 
\cite{Ma:2006km,Barbieri:2006dq,LopezHonorez:2006gr}. 

As mentioned above, CP can be spontaneously violated in models 
with three Higgs doublets with 
$Z_2$ symmetries \cite{Branco:1980sz}. 
Such is also the case for an $S_3$ symmetry \cite{Barradas-Guevara:2015rea,
Emmanuel-Costa:2016vej,Barradas-Guevara:2016ecp}.  

This talk is based on the work done in Ref.~\cite{Emmanuel-Costa:2016vej} where 
the possible vacuum solutions were analysed and the possibility of
having spontaneous CP violation in the context of three Higgs doublets,
with an $S_3$ symmetry, was studied. In this work we concentrated our attention 
on the scalar potential. Several authors have considered implications
of three-Higgs-doublet models with an $S_3$ symmetry for flavour physics
(see, for example: \cite{Pakvasa:1977in,Ma:1999hh,Araki:2005ec,
Bhattacharyya:2010hp, Teshima:2012cg, Canales:2013ura, Das:2015sca,Cruz:2017add}).

\section{Vacua of $S_3$-symmetric three-Higgs-doublet potential}
The  $S_3$ group is the permutation group involving three Higgs doublets 
$\phi_1$, $\phi_2$ and $\phi_3$, identified as the defining representation and
constituting a reducible triplet of Higgs doublets.
The study of the $S_3$-symmetric three-Higgs-doublet potential can be done
in two different frameworks, either in terms of the defining representation
\cite{Derman:1978rx, Derman:1979nf},
or in terms of the irreducible representations \cite{Kubo:2004ps}. 
In what follows, we classify the vacua in terms of constraints on 
the potential. Vacua can also be classified in terms of their residual symmetries \cite{Ivanov:2014doa}.

\subsection{The scalar potential}
The $S_3$ 
symmetric potential has a quadratic and a quartic part, which in terms 
of the defining representation can be written \cite{Derman:1978rx}:
\begin{equation}
V=V_2+V_4
\end{equation}
\begin{subequations} 
\begin{align}
\label{Eq:pot-original}
V_2 & =-\lambda\sum_{i}\phi_i^\dagger\phi_i +
\frac{1}{2} \gamma\sum_{i<j}[\phi_i^\dagger\phi_j+ \mbox{h.c.}], \\
V_4 & =  A\sum_{i}(\phi_i^\dagger\phi_i)^2
+\sum_{i<j}\{C(\phi_i^\dagger\phi_i)(\phi_j^\dagger\phi_j)
+\overline C (\phi_i^\dagger\phi_j)(\phi_j^\dagger\phi_i) 
+\frac{1}{2} D[(\phi_i^\dagger\phi_j)^2+\mbox{h.c.}]\} \nonumber \\
& +  \frac{1}{2} E_1\sum_{i\neq j}[(\phi_i^\dagger\phi_i)(\phi_i^\dagger\phi_j)+\mbox{h.c.}]
+\sum_{i\neq j\neq k\neq i,j<k}
\{\frac{1}{2} E_2[(\phi_i^\dagger\phi_j)(\phi_k^\dagger\phi_i)+\mbox{h.c.}] \quad 
\nonumber \\
& +   \frac{1}{2} E_3[(\phi_i^\dagger\phi_i)(\phi_k^\dagger\phi_j)+\mbox{h.c.}] 
+\frac{1}{2} E_4[(\phi_i^\dagger\phi_j)(\phi_i^\dagger\phi_k)+\mbox{h.c.}]\}.
\label{29ab}
\end{align}
\end{subequations}
Here all fields appear on an equal footing. This representation is not 
irreducible, it splits into two irreducible representations consisting of
a singlet, $h_S$ and a doublet of $S_3$ with components  $h_1$ and $h_2$.
The decomposition into these two irreducible representations is given by:
\begin{equation} 
\left(
\begin{array}{c}
h_1\\ h_2 \\h_S
\end{array}
\right)
=\left(
\begin{array}{ccc}
\frac{1}{\sqrt2} & - \frac{1}{\sqrt2} & 0 \\
\frac{1}{\sqrt6} & \frac{1}{\sqrt6} & -\frac{2}{\sqrt6} \\
\frac{1}{\sqrt3} & \frac{1}{\sqrt3} & \frac{1}{\sqrt3}
\end{array}
\right)
\left(
\begin{array}{c}
\phi_1 \\ \phi_2 \\ \phi_3
\end{array}
\right)
\end{equation}
This matrix has a striking similarity with the tribimaximal mixing 
matrix \cite{Harrison:2002er}
which is very close to the observed leptonic mixing.

The scalar potential written in term of fields from irreducible
representations has the form \cite{Kubo:2004ps, Teshima:2012cg,Das:2014fea}
\begin{subequations} \label{Eq:V-DasDey}
\begin{align}
V_2&=\mu_0^2 h_S^\dagger h_S +\mu_1^2(h_1^\dagger h_1 + h_2^\dagger h_2), \\
V_4&=
\lambda_1(h_1^\dagger h_1 + h_2^\dagger h_2)^2 
+\lambda_2(h_1^\dagger h_2 - h_2^\dagger h_1)^2
+\lambda_3[(h_1^\dagger h_1 - h_2^\dagger h_2)^2+(h_1^\dagger h_2 + h_2^\dagger h_1)^2]
\nonumber \\
&+ \lambda_4[(h_S^\dagger h_1)(h_1^\dagger h_2+h_2^\dagger h_1)
+(h_S^\dagger h_2)(h_1^\dagger h_1-h_2^\dagger h_2)+\hc] 
+\lambda_5(h_S^\dagger h_S)(h_1^\dagger h_1 + h_2^\dagger h_2) \nonumber \\
&+\lambda_6[(h_S^\dagger h_1)(h_1^\dagger h_S)+(h_S^\dagger h_2)(h_2^\dagger h_S)] 
+\lambda_7[(h_S^\dagger h_1)(h_S^\dagger h_1) + (h_S^\dagger h_2)(h_S^\dagger h_2) +\hc]
\nonumber \\
&+\lambda_8(h_S^\dagger h_S)^2.
\label{Eq:V-DasDey-quartic}
\end{align}
\end{subequations}
In this form the potential has no symmetry for the interchange of $h_1$ and 
$h_2$ but there is a $Z_2$ symmetry of the form $h_1 \rightarrow - h_1$.
There is an equivalent doublet representation which has also been used
in the literature \cite{Bhattacharyya:2010hp}:
\begin{eqnarray}
\left( \begin{array}{c} \hat\chi_1 \\ \hat\chi_2 
\end{array} \right)
= \frac{1}{\sqrt{2}}\left( \begin{array}{cc}  
i  & 1  \\
-i & 1   
\end{array}\right)
\left( \begin{array}{c} h_1 \\ h_2
\end{array} \right),
\end{eqnarray}
where the  symmetry appears as a symmetry for the interchange of
the fields $\hat\chi_1$ and $\hat\chi_2$. 
Both expressions for the potential describe the same physics and the coefficients
in the different frameworks are related through linear equations. 

With the special choice
of  $\lambda_4=0$ the potential acquires a continuous $SO(2)$ symmetry defined by:
\begin{eqnarray}
\left( \begin{array}{c} h_1^\prime \\ h_2^\prime
\end{array} \right)
= \left( \begin{array}{cc}
\cos \theta  & - \sin \theta  \\
\sin \theta & \cos \theta
\end{array}\right)
\left( \begin{array}{c} h_1 \\ h_2
\end{array} \right).
\label{so2}
\end{eqnarray}
Spontaneous breaking of this symmetry leads to a massless scalar which of
course must be avoided.

$S_3$ has three irreducible representations, a doublet, a singlet and a pseudosinglet, 
$h_A$. The latter has no direct translation into the initial fields 
$\phi_1$, $\phi_2$ and $\phi_3$ and transforms under $S_3$ into (-$h_A$).
From a group theoretical point of view we can 
choose to write an $S_3$ symmetric potential in terms of the doublet and
$h_A$. The new potential becomes:
\begin{subequations}
\begin{align}
V_2&=\mu_0^2 h_A^\dagger h_A +\mu_1^2(h_1^\dagger h_1 + h_2^\dagger h_2), \\
V_4&=
\lambda_1(h_1^\dagger h_1 + h_2^\dagger h_2)^2 
+\lambda_2(h_1^\dagger h_2 - h_2^\dagger h_1)^2
+\lambda_3[(h_1^\dagger h_1 - h_2^\dagger h_2)^2+(h_1^\dagger h_2 + h_2^\dagger h_1)^2]
\nonumber \\
&+ \lambda_4[(h_A^\dagger h_2)(h_1^\dagger h_2+h_2^\dagger h_1)
-(h_A^\dagger h_1)(h_1^\dagger h_1-h_2^\dagger h_2)+\hc] 
+\lambda_5(h_A^\dagger h_A)(h_1^\dagger h_1 + h_2^\dagger h_2) \nonumber \\
&+\lambda_6[(h_A^\dagger h_1)(h_1^\dagger h_A)+(h_A^\dagger h_2)(h_2^\dagger h_A)] 
+\lambda_7[(h_A^\dagger h_1)(h_A^\dagger h_1) + (h_A^\dagger h_2)(h_A^\dagger h_2) +\hc]
\nonumber \\
&+\lambda_8(h_A^\dagger h_A)^2,
\label{Eq:V-David-quartic}
\end{align}
\end{subequations}
which reduces to the same potential we had before with $h_1$ and $h_2$
interchanged. At this stage there is no new physics from this choice of 
representations. However this may change depending on how the couplings
to the fermions are introduced.

In order to study the possibility of having spontaneous CP violation
we start with a potential with real coefficients. This choice guarantees,
without loss of generality, that the potential conserves CP. 
In this case we are left with ten independent parameters irrespective 
of the choice of representations. This potential does not fall into
a CP conserving potential with irremovable complex parameters \cite{Ivanov:2015mwl}.

We use the following field notations for the decomposition of the $SU(2)$ Higgs
doublets:
\begin{equation} 
\phi_i=\left(
\begin{array}{c}\varphi_i^+\\ (\rho_i+\eta_i+i\chi_i)/\sqrt{2}
\end{array}\right), \quad i=1,2,3,
\label{Obasis}
\end{equation}

\begin{equation} \label{Eq:hi_hS}
h_i=\left(
\begin{array}{c}h_i^+\\ (w_i+\tilde \eta_i+i\tilde \chi_i)/\sqrt{2}
\end{array}\right), \quad i=1,2, \quad
h_S=\left(
\begin{array}{c}h_S^+\\ (w_S+\tilde \eta_S+i\tilde \chi_S)/\sqrt{2}
\end{array}\right).
\end{equation}

%%%%%%%%%%%%%%%%%%%%%%%%%%%%%%%%%%%%%%%%%%%%%
\begin{table}[htb]
\caption{Possible real vacua (partly after Derman and Tsao \cite{Derman:1979nf}).
This classification uses the notation R-X-y,
where R refers to ``real". The roman numeral X gives the number of constraints on the parameters of the potential that arise from solving the stationary-point equations. The letter y is used to distinguish  different vev's that have the same X, and 
$\lambda_a$ is defined in Eq.~(\ref{Eq:lambda_ab}).}
\label{Table:real}
\begin{center}
\begin{tabular}{|cccc|}
\hline
\hline
Vacuum  & $\rho_1,\rho_2,\rho_3$ & $w_1,w_2,w_S$ & Comment  \\
\hline
\hline
R-0& $0,0,0$  & $0,0,0$ & Not interesting  \\
\hline
\hline
R-I-1 & $x,x,x$  & $0,0,w_S$ & $\mu _0^2= -\lambda _8 w_S^2$  \\
\hline
R-I-2a & $x,-x,0$ & $w,0,0$ &  $\mu _1^2=- \left(\lambda _1+\lambda _3\right) w_1^2$\\
\hline
R-I-2b & $x,0,-x$ & $w, \sqrt{3} w, 0$ & $\mu _1^2= -\frac{4}{3} \left(\lambda _1+\lambda _3\right) w_2^2$ \\
\hline
R-I-2c & $0,x,-x$ & $w, -\sqrt{3} w, 0$ &$\mu _1^2= -\frac{4}{3}\left(\lambda _1+\lambda _3\right) w_2^2$  \\
\hline
\hline
R-II-1a & $x,x,y$ & $0,w,w_S$ & $\mu _0^2= \frac{1}{2}\lambda _4\frac{ w_2^3}{w_S}
-\frac{1}{2} \lambda_a w_2^2-\lambda _8 w_S^2$, \\
& & & $\mu _1^2= -\left( \lambda _1+ \lambda _3\right) w_2^2+\frac{3}{2} \lambda _4 w_2 w_S-\frac{1}{2} \lambda_a w_S^2$\\
\hline
R-II-1b & $x,y,x$ & $w,-w/\sqrt{3},w_S$ & $\mu _0^2= -4\lambda _4\frac{ w_2^3}{w_S}-2\lambda_a w_2^2-\lambda _8 w_S^2$, \\
& & & $\mu _1^2=-4 \left(\lambda _1+\lambda _3\right) w_2^2 -3 \lambda _4 w_2 w_S-\frac{1}{2} \lambda_a w_S^2$\\
\hline
R-II-1c & $y,x,x$ & $w,w/\sqrt{3},w_S$ &  $\mu _0^2= -4\lambda _4\frac{w_2^3}{w_S}-2\lambda_a w_2^2-\lambda _8 w_S^2$, \\
& & & $\mu _1^2= -4 \left(\lambda _1+\lambda _3\right) w_2^2-3 \lambda _4 w_2 w_S-\frac{1}{2} \lambda_a w_S^2$\\
\hline
R-II-2 & $x,x,-2x$ & $0, w, 0$ &$\mu _1^2= - \left(\lambda _1+\lambda _3\right) w_2^2$, $\lambda_4=0$  \\
\hline
R-II-3 & $x,y,-x-y$ & $w_1,w_2,0$ & $\mu _1^2= -\left(\lambda _1+\lambda _3\right)(w_1^2+w_2^2),\lambda_4=0$ \\
\hline
\hline
R-III & $\rho_1,\rho_2,\rho_3$ & $w_1,w_2,w_S$ & $\mu _0^2= -\frac{1}{2} \lambda_a( w_1^2+ w_2^2)-\lambda _8 w_S^2$,\\
& & & $\mu _1^2= - \left( \lambda _1+ \lambda _3\right)( w_1^2+ w_2^2)-\frac{1}{2} \lambda_a w_S^2$, \\
& & & $\lambda_4=0$ \\
\hline
\hline
\end{tabular}
\end{center}
\end{table}
%%%%%%%%%%%%%%%%%%%%%%%%%%%%%%%%%%%%%%%%%%%%%

\subsection{Real vacuum solutions}
Real vacuum solutions do not violate CP spontaneously. It is interesting to
understand what are the possible real solutions for the vacuum. In this case 
one has to solve three minimisation conditions corresponding to the vanishing
of the three relevant derivatives of the potential. In the irreducible framework
these conditions can be solved in terms of $\mu_0^2$ and $\mu_1^2$ 
leading to \cite{Das:2014fea}:
\begin{subequations} \label{Eq:DD-mu0-mu1}
\begin{align} \label{Eq:mu_0_sq}
\mu_0^2&=\frac{1}{2w_S}
\left[ \lambda_4(w_2^2-3w_1^2)w_2 -(\lambda_5+\lambda_6+2\lambda_7)(w_1^2+w_2^2)w_S -2\lambda_8w_S^3
\right], \\
\mu_1^2&=-\frac{1}{2}
\left[2(\lambda_1+\lambda_3) (w_1^2+w_2^2)+6\lambda_4w_2w_S +(\lambda_5+\lambda_6+2\lambda_7)w_S^2
\right], \label{Eq:DD-mu_1-a}\\
\mu_1^2&=-\frac{1}{2}
\left[2(\lambda_1+\lambda_3) (w_1^2+w_2^2)
-3\lambda_4(w_2^2-w_1^2)\frac{w_S}{w_2}
+(\lambda_5+\lambda_6+2\lambda_7)w_S^2 
\right]. \label{Eq:DD-mu_1-b}
\end{align}
\end{subequations}
The first equation comes from the derivative of the potential with respect to $w_S$
and the second and third from the derivatives with respect to $w_1$ and $w_2$.
Eqs.~(\ref{Eq:DD-mu_1-a}) and (\ref{Eq:DD-mu_1-b}) were obtained dividing by 
$w_1$ and $w_2$ respectively. Clearly, these two equations 
are not automatically consistent. There are several possible consistency cases:
\begin{itemize}
\item
for $w_1=0$ the corresponding derivative is zero and there is no clash
with the determination of $\mu_1^2$ from Eq.~(\ref{Eq:DD-mu_1-b}).
\item
otherwise, $\lambda_4(3w_2^2-w_1^2)w_S=0$ is required. This can be achieved in 
three different ways: $\lambda_4=0$ or $w_1=\pm\sqrt{3}w_2$ or $w_S=0$.
\item
for $\lambda_4=0$ a special condition arises from Eq.~(\ref{Eq:mu_0_sq}):
$\lambda_4 w_2(3w_1^2-w_2^2)=0$ so that in addition we must have 
$\lambda_4=0$ or $w_2=\pm\sqrt{3}w_1$, or  $w_2=0$.
\end{itemize}
Derman and Tsao \cite{Derman:1979nf} analysed spontaneous symmetry breaking 
with real vacua taking also into account the residual symmetries. Their work was 
done in the reducible framework where the condition $\lambda_4=0$
corresponds to $4A-2(C+\overline C+D)-E_1+E_2+E_3+E_4=0$. This
condition was obtained before by Derman \cite{Derman:1978rx} who considered
it very unnatural, since in his context it was not clear that it
was associated to an additional symmetry. With $\lambda_4 \neq 0$ 
there were only three possible real solutions \cite{Derman:1979nf}:
\begin{itemize}
\item
$(x,x,x)$ leaving $S_3$ unbroken  and translating into the 
doublet-singlet notation as $(0, 0, w_S)$; consistency condition: $w_1 = 0$
(also verifies  $w_1=\pm\sqrt{3}w_2$).
\item
$(x,x,y)$ leaving a residual $S_2$ symmetry. In terms of the reducible 
representation any ordering of the vevs is equivalent, however, in
the definition of the doublet of $S_3$ a special direction is chosen. As
a result, different orderings correspond to different translations: \\
$(x,x,y)$ translates into $(0, w_2, w_S)$; consistency condition: $w_1=0$. \\
$(x,y,x)$ translates into $(w_1, - \frac{1}{\sqrt{3}} w_1 , w_S)$; consistency condition:
$w_1= - \sqrt{3}w_2$. \\
$(y,x,x)$ translates into $(w_1,  \frac{1}{\sqrt{3}} w_1 , w_S)$; consistency condition:
$w_1=  \sqrt{3}w_2$.
\item
$(x, y, z) = (x, -x, 0)$ leaving a residual $S_2$ symmetry. This is the only 
possible real solution with all three vevs different from each other, unless one 
imposes $4A-2(C+\overline C+D)-E_1+E_2+E_3+E_4=0$ ($\lambda_4=0$). The translation
into the irreducible representation is now: \\
$(x, -x, 0)$ translates into $(w_1 = \sqrt{2} x , 0 , 0)$: consistency conditions: $w_S=0$ together with $w_2=0$. \\
$(x, 0, -x)$ translates into $(w_1 = \frac{1}{\sqrt{2}} x , w_2 = \frac{\sqrt{3}}
{\sqrt{2}} x, 0)$; consistency conditions: $w_S=0$ together with $w_2= \sqrt{3} w_1$. \\
$(0, x, -x)$ translates into $(w_1 = - \frac{1}{\sqrt{2}} x , w_2 = \frac{\sqrt{3}}
{\sqrt{2}} x, 0)$; consistency conditions: $w_S=0$ together with $w_2= - \sqrt{3} w_1$. 
\end{itemize}
Table 1 summarises all the possible real solutions together with the
constraints imposed on the parameters of the potential. The following 
abbreviation was introduced: 
\begin{equation} \label{Eq:lambda_ab}
\lambda_a=\lambda_5+\lambda_6+2\lambda_7.
\end{equation}

%%%%%%%%%%%%%%%%%%%%%%%%%%%%%%%%%%%%%%%%%%%%%%%%%%%%%%%%%%%%%%%%
%\begin{figure}[htb]
%\refstepcounter{figure}
%\label{Fig:br-top-limit}
%\addtocounter{figure}{-1}
%\begin{center}
%\includegraphics[scale=0.10]{DSC0025}
%\end{center}
%\vspace*{-4mm}
%\caption{Happy author/speaker.}
%\end{figure}
%%%%%%%%%%%%%%%%%%%%%%%%%%%%%%%%%%%%%%%%%%%%%%%%%%%%%%%%%%%%%%%%

%%%%%%%%%%%%%%%%%%%%%%%%%%%%%%%%%%%%%%%%%%%%%
\begin{table}[htb]
\caption{Complex vacua. Notation: $\epsilon=1$ and $-1$ for C-III-d and C-III-e, respectively;
$\xi=\sqrt{-3\sin 2\rho_1/\sin2\rho_2}$,
$\psi=\sqrt{[3+3\cos (\rho_2-2 \rho _1)]/(2\cos\rho_2)}$. Due to the constraints 
the vacua labelled with an asterisk ($^\ast$) are in fact real.}
\label{Table:complex}
\begin{center}
\begin{tabular}{|ccc|}
\hline\hline
& IRF (Irreducible Rep.)& RRF  (Reducible Rep.) \\
\hline
& $w_1,w_2,w_S$ & $\rho_1,\rho_2,\rho_3$  \\
\hline
\hline
C-I-a & $\hat w_1,\pm i\hat w_1,0$ & 
$x, xe^{\pm\frac{2\pi i}{3}}, xe^{\mp\frac{2\pi i}{3}}$ \\
\hline
\hline
C-III-a & $0,\hat w_2e^{i\sigma_2},\hat w_S$ & $y, y, xe^{i\tau}$  \\
\hline
C-III-b & $\pm i\hat w_1,0,\hat w_S$ & $x+iy,x-iy,x$  \\
\hline
C-III-c & $\hat w_1 e^{i\sigma_1},\hat w_2e^{i\sigma_2},0$ 
& $xe^{i\rho}-\frac{y}{2}, -xe^{i\rho}-\frac{y}{2}, y$  \\
\hline
C-III-d,e & $\pm i \hat w_1,\epsilon\hat w_2,\hat{w}_S$ & $xe^{ i\tau},xe^{- i\tau},y$ \\
\hline
C-III-f & $\pm i\hat w_1 ,i\hat w_2,\hat{w}_S$ 
& $re^{i\rho}\pm ix,re^{i\rho}\mp ix,\frac{3}{2}re^{-i\rho}-\frac{1}{2}re^{i\rho}$ \\
\hline
C-III-g & $\pm i\hat w_1,-i\hat w_2,\hat{w}_S$ 
& $re^{-i\rho}\pm ix,re^{-i\rho}\mp ix,\frac{3}{2}re^{i\rho}-\frac{1}{2}re^{-i\rho}$ \\
\hline
C-III-h & $\sqrt{3}\hat w_2 e^{i\sigma_2},\pm\hat w_2 e^{i\sigma_2},\hat{w}_S$ 
& $xe^{i\tau} , y , y$ \\
& & $y, xe^{i\tau},y$ \\
\hline
C-III-i & $\sqrt{\frac{3(1+\tan^2\sigma_1)}{1+9\tan^2\sigma_1}}\hat w_2e^{i\sigma_1},$ 
& $x, ye^{i\tau},ye^{-i\tau}$ \\
& $\pm\hat w_2e^{-i\arctan(3\tan\sigma_1)},\hat w_S$ 
& $ye^{i\tau}, x, ye^{-i\tau}$ \\
\hline
\hline
C-IV-a$^\ast$ & $\hat w_1e^{i\sigma_1},0,\hat w_S$ & $re^{i\rho}+x, -re^{i\rho}+x,x$ \\
\hline
C-IV-b & $\hat w_1,\pm i\hat w_2,\hat w_S$ 
& $re^{i\rho}+x, -re^{-i\rho}+x, -re^{i\rho}+re^{-i\rho}+x$ \\
\hline
C-IV-c & $\sqrt{1+2\cos^2\sigma_2}\hat w_2,$ &  $re^{i\rho}+r\sqrt{3(1+2\cos^2\rho)}+x$, \\
& $\hat w_2e^{i\sigma_2},\hat w_S$ 
& $re^{i\rho}-r\sqrt{3(1+2\cos^2\rho)}+x,-2re^{i\rho}+x$ \\
\hline
C-IV-d$^\ast$ & $\hat w_1e^{i\sigma_1},\pm\hat w_2e^{i\sigma_1},\hat w_S$ & $r_1e^{i\rho}+x, (r_2-r_1)e^{i\rho}+x,-r_2e^{i\rho}+x$ \\
\hline
C-IV-e & $\sqrt{-\frac{\sin 2\sigma_2}{\sin 2\sigma_1}}\hat w_2e^{i\sigma_1},$ & $re^{i\rho_2}+re^{i\rho_1}\xi+x, re^{i\rho_2}-re^{i\rho_1}\xi+x,$  \\
& $\hat w_2e^{i\sigma_2},\hat w_S$ & $-2re^{i\rho_2}+x$ \\
\hline
C-IV-f & $\sqrt{2+\frac{\cos \left(\sigma _1-2 \sigma _2\right)}{\cos\sigma_1}}\hat w_2e^{i\sigma_1},$ & $re^{i\rho_1}+re^{i\rho_2}\psi+x$, \\
& $\hat w_2e^{i\sigma_2},\hat w_S$ &$re^{i\rho_1}-re^{i\rho_2}\psi+x, -2re^{i\rho_1}+x$ \\
\hline
\hline
C-V$^\ast$ & $\hat w_1e^{i\sigma_1},\hat w_2e^{i\sigma_2},\hat w_S$ & $xe^{i\tau_1},ye^{i\tau_2},z$ \\
\hline
\end{tabular}
\end{center}
\end{table}
%%%%%%%%%%%%%%%%%%%%%%%%%%%%%%%%%%%%%%%%%%%%%

\subsection{Complex vacuum solutions}
In the discussion of possible complex vacua we now adopt a convention
where $w_S$ is real and non-negative and take 
\begin{equation} \label{Eq:IRF:notation}
w_1=\hat w_1e^{i\sigma_1}, \quad 
w_2=\hat w_2e^{i\sigma_2},
\end{equation}
with the $\hat {w_i}$ also real and non-negative. With this convention 
$w_S$ is also denoted by $\hat{w_S}$. A systematic analysis of possible solutions
was performed in \cite{Emmanuel-Costa:2016vej}. The results are summarised 
in Table 2. The list of the constraints on the potential
that are consistent with each solution is not given here, it can be found 
in Ref.~\cite{Emmanuel-Costa:2016vej}.

Several solutions require $\lambda_4=0$. This is not a new feature, it also
happened in the context of real solutions. For $\lambda_4=0$
the potential acquires a continuous $SO(2)$ symmetry which can be broken
spontaneously by the vacuum solutions, therefore, leading to a 
massless scalar. Massless scalars are ruled out by experiment. It is
possible to avoid this problem by introducing soft breaking terms.
The most general form for the $V_2$ part of the potential with soft breaking terms
would be:
\begin{align}
V_2 =  \mu_0^2 & h_S^\dagger h_S +\mu_1^2(h_1^\dagger h_1 + h_2^\dagger h_2) +
\mu_2^2(h_1^\dagger h_1 - h_2^\dagger h_2) + \frac{1}{2} \nu^2 
(h_2^\dagger h_1 - h_1^\dagger h_2) \nonumber \\
+ \mu_3^2 & (h_S^\dagger h_1 - h_1^\dagger h_S) +
\mu_4^2 (h_S^\dagger h_2 - h_2^\dagger h_S).
\end{align}
However, soft breaking terms involving $h_S$ and one $h_i$ are not consistent with
$\lambda_4=0$.

In Table 3 we collect all possible complex vacuum solutions indicating whether or 
not they require $\lambda_4$ equal to zero and whether or not they allow for 
spontaneous CP violation. One important conclusion from our analysis is that
there are cases where CP can be violated spontaneously, however, 
no solution requiring  $\lambda_4=0$ can lead to spontaneous CP violation.
In order to confirm that CP could indeed be violated spontaneously
we used a powerful tool based on CP-odd Higgs-basis-invariant conditions,
verifying that there were indeed conditions that were violated.
There are several such conditions which were especially built for the
analysis of the Higgs potential
\cite{Lavoura:1994fv,Branco:2005em, Gunion:2005ja,Varzielas:2016zjc,Branco:1999fs}. 
In the next subsection we discuss 
spontaneous CP violation using a few illustrative 
examples.

%%%%%%%%%%%%%%%%%%%%%%%%%%%%%%%%%%%%%%%%%%%%%
\begin{table}[htb]
\caption{Spontaneous CP violation}
\label{Table:CPV}
\begin{center}
\begin{tabular}{|ccc||ccc||ccc|}
\hline
Vacuum  &  $\lambda_4$ & SCPV & Vacuum  &  $\lambda_4$ &  SCPV 
& Vacuum &  $\lambda_4$ &  SCPV\\
\hline
C-I-a & X & no & C-III-f,g & 0 & no & C-IV-c & X & yes   \\
C-III-a & X & yes  & C-III-h & X & yes  & C-IV-d & 0 & no \\
C-III-b & 0 & no  & C-III-i & X & no & C-IV-e & 0 & no \\
C-III-c & 0 & no  & C-IV-a & 0 & no & C-IV-f & X & yes \\
C-III-d,e & X & no  & C-IV-b & 0 & no & C-V & 0 & no\\
\hline
\end{tabular}
\end{center}
\end{table}
%%%%%%%%%%%%%%%%%%%%%%%%%%%%%%%%%%%%%%%%%%%%%

\subsection{Spontaneous CP violation}
Spontaneous CP violation can only occur if the Lagrangian conserves CP but the 
vacuum does not. This can only happen when there is no transformation that can be 
identified with a CP transformation leaving both the Lagrangian and the vacuum 
invariant. Under a CP transformation a single Higgs doublet $\Phi$ 
transforms into its complex conjugate. In models with several Higgs doublets
the most general CP transformation is given by:
\begin{equation}
\Phi_i \overset{\mbox{CP}}{\longrightarrow} U_{ij} \Phi^\ast_j.
\end{equation}
Here, $U$ is a unitary matrix mixing different Higgs doublets and corresponds to a Higgs
basis transformation\footnote{This transformation is often referred to
as a ``generalized" CP transformation, thus suggesting that there is 
also a ``non-generalized" CP transformation. This is, of course, misleading.}. 
Higgs basis transformations do not change the physics.

If all the coefficients of the potential are real the potential conserves CP
explicitly and the above equation is verified for $U$ the identity matrix. 
Checking for explicit CP invariance of a multi-Higgs potential may be
a non-trivial task since Higgs basis transformations, in general, can transform
couplings that are real in one Higgs basis into couplings that are complex in
another basis. For this purpose CP-odd Higgs basis invariants are of great help 
\cite{Branco:2005em, Gunion:2005ja}. Once it is known that a Lagrangian
conserves CP it remains to check whether or not CP is violated spontaneously. 
It has been shown \cite{Branco:1983tn} that in order for the vacuum to 
conserve CP the following relation has to be obeyed: 
\begin{equation}
 U_{ij} \langle 0| \Phi_j |0\rangle^\ast = \langle 0| \Phi_i |0\rangle
\label{geo}
\end{equation}
with $U$ now a unitary matrix corresponding to a symmetry of the Lagrangian.
This relation is very powerful and allows to show that vacua that 
would at first sight violate CP are indeed CP conserving. This can be 
illustrated with a few examples taken from Table 3. For a full discussion see
Ref.~\cite{Emmanuel-Costa:2016vej}.

\begin{itemize}
\item
Let us consider the vacuum identified as C-I-a, given by 
$\left( x, xe^{\pm\frac{2\pi i}{3}}, xe^{\mp\frac{2\pi i}{3}} \right) $ 
in the reducible representation. It is not possible to rephase the 
three Higgs doublets in such a way that the three vevs become real 
keeping at the same time the potential real. This is a vacuum 
solution with calculable non-trivial phases, fixed by the symmetry of
the potential with no explicit dependence on the parameters of the 
potential. Such phases are called geometrical phases \cite{Branco:1983tn}.
It was shown in Ref.~\cite{Branco:1983tn} that this vacuum does not violate CP
since Eq.~(\ref{geo}) can be verified for $U$ given by:
\begin{equation}
U = \left( \begin{array}{ccc}
1 & 0 & 0 \\
0 & 0 & 1 \\
0 & 1 & 0 
\end{array} \right).
\end{equation}
This matrix makes use of the symmetry of the potential for the interchange
of $\phi_2$ and $\phi_3$.
\item
Another interesting example is the C-III-c vacuum which is of the form
$\left( \hat w_1 e^{i\sigma_1},\hat w_2e^{i\sigma_2},0 \right) $ 
in the irreducible representation framework. It can also be written, 
without loss of generality, through an overall phase rotation,
in the form  $\left( \hat w_1 e^{i\sigma},\hat w_2,0 \right) $. At first sight
this vacuum looks like a CP violating vacuum, especially taking into
consideration the fact that the moduli of $w_1$ and $w_2$ are different
from each other. However, once again we can use Eq.~(\ref{geo}) to show that 
this vacuum conserves CP. Notice that this solution requires $\lambda_4 = 0$
(see Table 3)
and therefore there is an $SO(2)$ symmetry for the fields $h_1$ and $h_2$.
With this knowledge one can build the necessary matrix $U$ and Eq.~(\ref{geo})
becomes:
\begin{equation}
e^{i(\delta_1+\delta_2)}
\begin{pmatrix}
\cos\theta &\sin\theta & 0\\
-\sin\theta & \cos\theta & 0 \\
0 & 0 & 1
\end{pmatrix}
\begin{pmatrix}
0 & 1 & 0 \\
1 & 0 & 0 \\
0 & 0 & 1
\end{pmatrix}
\begin{pmatrix}
\cos\theta &-\sin\theta & 0 \\
\sin\theta & \cos\theta & 0 \\
0 & 0 & 1
\end{pmatrix}
\begin{pmatrix}
\hat w_1e^{i\sigma} \\
\hat w_2 \\
0
\end{pmatrix}^\ast
=
\begin{pmatrix}
\hat w_1e^{i\sigma} \\
\hat w_2 \\
0
\end{pmatrix},
\label{eq10}
\end{equation}
or
\begin{equation}
e^{i(\delta_1+\delta_2)}
\begin{pmatrix}
\sin2\theta &\cos2\theta & 0\\
\cos2\theta & -\sin2\theta & 0 \\
0 & 0 & 1
\end{pmatrix}
\begin{pmatrix}
\hat w_1e^{i\sigma} \\
\hat w_2 \\
0
\end{pmatrix}^\ast
=
\begin{pmatrix}
\hat w_1e^{i\sigma} \\
\hat w_2 \\
0
\end{pmatrix}.
\end{equation}
In this example the matrix $U$ has several components: \\
- an $SO(2)$ rotation of $h_1$ and $h_2$ by an angle 
$\theta$, which should be chosen as:
\begin{equation}
\tan 2 \theta = \frac{\hat w_1^2-\hat w_2^2}{2\hat w_1\hat w_2 \cos\sigma}.
\end{equation}
With this choice, the vevs of the new $S_3$ doublet fields acquire
the same modulus and the new vacuum acquires the form $(ae^{i\delta_1},ae^{i\delta_2},0)$, \\
- an overall phase rotation of the three Higgs doublets by 
$\exp[-i(\delta_1+\delta_2)/2]$, so that now the first two vevs acquire symmetric phases: 
$(ae^{i\delta},ae^{-i\delta},0)$, \\
- finally we just need to use the symmetry for the interchange 
$h_1^\prime \leftrightarrow h_2^\prime$  in the $S_3$ doublet representation.
\end{itemize}
The last example illustrates how powerful the condition given by Eq.~(\ref{geo}) can be,
but at the same time it shows that, as complexity grows,
it may be non-trivial, in cases where such a matrix exists, 
to build the necessary matrix $U$.
In fact this may require special insight and there is the danger of
missing it, in a CP conserving case. In Ref.~\cite{Ogreid:2017alh}   
we propose an alternative simple 
method, which is very useful in such cases, and allows to detect or eliminate 
the possibility of having spontaneous CP violation in 
multi-Higgs models. The three tools, consisting of the use of
CP-odd invariant conditions, the relation given by Eq.~(\ref{geo}) and
the simple method proposed in Ref.~\cite{Ogreid:2017alh}, combined
together, provide a reliable procedure to determine whether or not
a given Higgs potential violates CP spontaneously.

\section{Conclusions} 
We have presented here a summary of the work done in 
Ref.~\cite{Emmanuel-Costa:2016vej}.
We have focused on some important features of three-Higgs-doublet
models with an $S_3$ symmetry with emphasis on the discussion of
spontaneous CP violation.  Some aspects which were dealt with in the
paper were not included in this short presentation. We refer the
reader to the original work for a more detailed discussion 
of these aspects and for 
other topics such as ideas about constraining the potential
by the vevs, relations among complex and real vacua, and a discussion on
positivity beyond the necessary conditions given by 
Das and Dey \cite{Das:2014fea}
following the approach of Refs.~\cite{ElKaffas:2006nt,Grzadkowski:2009bt}
(see also \cite{EmmanuelCosta:2007zz}).   
Models with multi-Higgs doublets such as those discussed in our work
are very interesting and can 
in principle provide answers for several open questions. In particular 
they can provide viable dark matter candidates. 
These and other questions such as ways of generating realistic fermion masses
and mixing in this context or 
looking for viable models with spontaneous CP violation are still challenging
despite the fact that a lot of work has been already done
along these lines. These questions are very timely due to the
potential for being tested at the LHC.

\section*{Acknowledgements}
MNR thanks the local organising committee of the Symposium
Discrete 2016 for the stimulating scientific atmosphere and the warm
hospitality in Warsaw, and acknowledges financial support from 
the National Science Centre, Poland, the HARMONIA project under 
contract UMO-2015/18/M/ST2/00518 (2016-2019)
to participate in the Symposium and in the 1st Harmonia Meeting that
took place immediately after the Symposium. Special thanks go to 
Maria Krawczyk.
The work of MNR  is partially supported by Funda\c c\~ ao para a Ci\^ encia
e a Tecnologia (FCT, Portugal) through the projects CERN/FIS-NUC/0010/2015,
CFTP-FCT Unit 777 (UID/FIS/00777/2013) which are partially funded through
POCTI (FEDER), COMPETE, QREN and EU.

\end{document}